# Definitions of distance function in radial basis function approach

**W. Chen**

*Department of Mechanical System Engineering, Shinshu University, Wakasato 4-17-1, Nagano City, Nagano 380-8533, Japan*
*(E-mail:* * *chenw@homer.shinshu-u.ac.jp and chenw6@hotmail.com)*

## 1. Introduction

Very few studies involve how to construct the efficient RBFs by means of problem features. Recently the present author presented general solution RBF (GS-RBF) methodology to create operator-dependent RBFs successfully [1]. On the other hand, the normal radial basis function (RBF) is defined via Euclidean space distance function or the geodesic distance [2]. This purpose of this note is to redefine distance function in conjunction with problem features, which include problem-dependent and time-space distance function.

## 2. Problems with varying parameter, time-dependent or nonlinear terms

The GS-RBF [1] is constructed based on the canonical form of some operators such as the known Laplace or Helmholtz operators. Many important cases in engineering do not possess such standard form of operators. This section will show that some cares may be taken to handle these problems.

### 2.1 Varying parameter problems

The general second order partial differential system with varying coefficient can be stated as

$$R\frac{\partial^2 u}{\partial x^2} + S\frac{\partial^2 u}{\partial x \partial y} + T\frac{\partial^2 u}{\partial y^2} = 0, \quad (1)$$

where $R$, $S$ and $T$ are continuous functions of x and y. We can translate it into the canonical form by a suitable change of independent variables [2]

$$\xi = f_1(x,y), \quad \eta = f_2(x,y), \quad (2)$$

The corresponding distance function of the Euclidean norm is given by

$$r = \sqrt{(\xi - \xi_j)^2 + (\eta - \eta_j)^2}. \quad (3)$$

Substituting Eq. (2) into Eq. (3), we define the distance function on the original independent variables of x and y as

$$r = \sqrt{\begin{aligned}&[f_1(x,y) - f_1(x_j,y_j)]^2\\&+[f_2(x,y) - f_2(x_j,y_j)]^2\end{aligned}} \quad (4)$$

Obviously, the above definition of the distance function in terms of variable x and y are different from the direct use of the Euclidean norm.

In the following we illustrate two important special cases. Let us consider [3]

$$y^m\frac{\partial^2 u}{\partial x^2} + \frac{\partial^2 u}{\partial y^2} = 0, \quad (y \geq 0, m \succ -2), \quad (5)$$

its general solution is

$$u = r_2^{-2\beta}w(r^2), \quad (6)$$

where $\beta = m/2(m+2)$, $w$ is the hypergeometric functions,

$$r = \frac{r_1}{r_2}, \quad (7a)$$

$$r_1 = \sqrt{(x - x_j)^2 + \frac{4}{(m+2)^2}\left(y^{\frac{m+2}{2}} - y_j^{\frac{m+2}{2}}\right)^2}, \quad (7b)$$

$$r_2 = \sqrt{(x - x_j)^2 + \frac{4}{(m+2)^2}\left(y^{\frac{m+2}{2}} + y_j^{\frac{m+2}{2}}\right)^2} \quad (7c)$$

Let

$$\xi = x, \qquad \eta = \frac{2}{m+2}y^{\frac{m+2}{2}}, \quad (8)$$

we have

$$\frac{\partial^2 u}{\partial \xi^2} + \frac{\partial^2 u}{\partial \eta^2} = 0. \quad (9)$$

Substituting Eq. (8) into Eq. (3) also yields the distance function Eq. (7b).

For another example, let us consider

$$\frac{\partial^2 u}{\partial x^2} + y\frac{\partial^2 u}{\partial y^2} + \alpha\frac{\partial u}{\partial y} = 0, \quad (y \geq 0), \quad (10)$$

where $\alpha$ is coefficients. Its solution is

$$u = r_2^{-2\beta} w(r^2), \quad (11)$$

where $w$ is also the hypergeometric function,

$$\beta' = 1 - \beta = 1/2 - \alpha, \quad (12a)$$

$$r_1 = \sqrt{(x - x_j)^2 + 4\left(y^{l/2} - y_j^{l/2}\right)^2}, \quad (12b)$$

$$r_2 = \sqrt{(x - x_j)^2 + 4\left(y^{l/2} + y_j^{l/2}\right)^2} \quad (12c)$$

$$r = \frac{r_1}{r_2}, \quad (12d)$$

By using the independent variable transformation approach as in the previous case, we can get the distance functions of the RBF as shown in Eqs. (12a-d).

This hints us to define the distance functions simultaneously for the RBF $\psi(r_1, r_2)$

$$\psi(r_1, r_2) = \varphi(r_2)\phi(r_1/r_2), \quad (13)$$

The above examples show that for one certain problem, we can use multiple distance variables and RBFs simultaneously rather than one distance function and one RBF.

## 2.2 Nonlinear problems

The GS-RBF approach is applicable to the nonlinear problems. It is usually not easy to get the fundamental solution of general nonlinear problems. In some cases, however, we can transfer the nonlinear operator into the linear one via a variable transformation [4]. For example, consider the stationary heat transfer through the nonlinear materials where the conductivity $K$ is often a function of temperature $u$. The governing equation is given by

$$\frac{\partial}{\partial x}\left(K\frac{\partial u}{\partial x}\right) + \frac{\partial}{\partial y}\left(K\frac{\partial u}{\partial y}\right) - v_x\frac{\partial u}{\partial x} -$$
$$v_y\frac{\partial u}{\partial y} - u = \rho c\frac{\partial u}{\partial t}, \quad (15)$$

where $u$ is the temperature, $v_x$ and $v_y$ are the components of the velocity vector $v$, $\rho$ and $c$ are density and the specific heat of the materials, respectively. The coefficient $K$ is time-dependent conductivity coefficient

$$K = K_0\left[1 + \beta\left(\frac{u - u_0}{u_0}\right)\right]. \quad (16)$$

If we intend to solve this problem using the domain-type RBF, one of the reasonable choices of the RBF should be the TPS-type RBF due to the presence of Laplace-like operator. However, the brutal use of the TPS is discouraged in this case. By using the Kirchhoff transformation [4],

$$U = \int_{u_0}^{u} K(u)du, \quad (17)$$

we have

$$U = \frac{K_0 u_0}{2\beta}\left[1 + \beta\left(\frac{u - u_0}{u_0}\right)\right]^2 - \frac{K_0 u_0}{2\beta}. \quad (18)$$

The nonlinear Laplace operator is transformed

to the linear Laplace one

$$\frac{\partial^2 U}{\partial x^2} + \frac{\partial^2 U}{\partial z^2} - V_x \frac{\partial U}{\partial x} -$$
$$V_y \frac{\partial U}{\partial y} - f(U) = g(U)\frac{\partial U}{\partial t}, \tag{19}$$

where $V_x$, $V_y$, $f(U)$ and $g(U)$ are the functions of $U$. The TPS RBF $r^{2m}\ln(r)$ is recommended to solve Eq. (19). However, our purpose is to directly solve Eq. (15). According to Eq. (18), we have

$$u = u_0 + \frac{u_0}{\beta}\left(1 + \frac{2\beta U}{K_0 u_0}\right)^{1/2} - \frac{u_0}{\beta}. \tag{20}$$

The corresponding TPS RBF for $u$ is

$$\psi(r) = u_0 + \frac{u_0}{\beta}\left(1 + \frac{2\beta r^{2m}\ln(r)}{K_0 u_0}\right)^{1/2} - \frac{u_0}{\beta}, \tag{21}$$

which should be used in the RBF collocation of the original governing equation (15).

From the above analysis, we conclude that the use of RBF should fully use the features of the problems.

## 2.3. Time-space RBFs

The free symmetrical vibration of a very large membrane are governed by the equation

$$\frac{\partial^2 z}{\partial r^2} + \frac{1}{r}\frac{\partial z}{\partial r} = \frac{1}{c^2}\frac{\partial^2 z}{\partial t^2} \tag{22}$$

with $z = f(r)$, $\partial z/\partial t = g(r)$ when $t = 0$. We have the solution

$$z(r,t) = \int_0^\infty \bar{\xi}\bar{f}(\xi)\cos(\xi ct)J_0(\xi r)d\xi + \frac{1}{c}\int_0^\infty \bar{g}(\xi)\sin(\xi ct)J_0(\xi r)d\xi \tag{23}$$

where upper-dashed $f$ and $g$ are the zero-order Hankel transforms of $f(r)$ and $g(r)$, respectively. According to the GS-RBF, we have

$$z(r,t) = \left[\cos(ct) + \frac{1}{c}\sin(ct)\right]J_0(r) \tag{24}$$

The present author introduced concept of time-space distance variable and corresponding RBFs for time-dependent problems [1]. Here we hope to further improve its definition. Consider the equation governing wave propagation

$$u_{xx} = \frac{1}{c^2}u_{tt} + f(x,t). \tag{25}$$

Let

$$s = ict, \tag{26}$$

where $i$ means unit imaginary number, we have

$$u_{xx} + u_{ss} = f(x,t). \tag{27}$$

By analogy with the Euclidean definition of distance variable, the time-space RBFs (TSR) is defined

$$r_j = \sqrt{\left(x - x_j\right)^2 + \left(s - s_j\right)^2}$$
$$= \sqrt{\left(x - x_j\right)^2 - c^2\left(t - t_j\right)^2}. \tag{28}$$

However, such definition can lead to imaginary number of distance variable. Thus, it is safe to use

$$r_j = \sqrt{\left(x - x_j\right)^2 + c^2\left(t - t_j\right)^2}. \tag{29}$$

The above definition of $r$ differs from the standard radial basis function in that the time variable is included into the distance function and is handled equally as the other space variables. We can construct the RBF by means of the GS-RBF in such a way that time-dependence is naturally eliminated. In general, hyperbolic and elliptic equations have solutions whose arguments have the form $x + at$ and $x + ibt$ respectively, where $a$ and $b$ are real. Namely,

$$u(x,t) = f(p) \tag{30}$$

where $p$ is some unknown function of $x$ and $t$. For 3D case,

$$p = lx + my + nz + \mu_l \quad (31)$$

This provides some theoretical support to use time-space distance functions (28) and (29). However, the above situations do not exist for diffusion equation

$$\nabla^2 u = \frac{1}{k} u_t + f(x,t). \quad (32)$$

Its fundamental solution is given by

$$u^* = \frac{1}{(t_j - t)^{d/2}} \exp\left(-\frac{r^2}{4k(t_j - t)}\right) H(t_j - t), \quad (33)$$

where $d$ is the space dimensionality, $H()$ is the Heaviside function. Therefore, we give the corresponding RBF

$$\phi(r) = h(r) u^*(r, t, t_j), \quad (34)$$

where $h(r)$ is chosen according to problem feature. It is stressed that the response and source nodes must be totally differentially placed to avoid singularity in this case. This strategy of node placement is called as response source points staggering.

On the other hand, Chen and Tanaka [5] proposed time-space non-singular general solution

$$u^* = A e^{-k(t - t_j)} \phi(r), \quad (35)$$

for diffusion problem and

$$u^* = \left[ C \cos\left(c(t - t_j)\right) + D \sin\left(c(t - t_j)\right) \right] \phi(r) \quad (36)$$

for wave problems, where $\phi(r)$ is the zero order Bessel function of the first kind for 2D problems and $\sin(r)/r$ for 3D problems. Substituting these non-singular general solution into Eq. (34) produces the time-space RBF without singularity.

In addition, it is stressed that the time-space distance function and corresponding RBFs are applicable to transient data processing such as motion picture and movie.

## References

1. Chen, W. and Tanaka, M., New Insights in Boundary-only and Domain-type RBF Methods, *Inter. J. Nonlinear Sci. Numer. Simulation*, **1**(2000), 145-152.
2. Golberg, M.A. and Chen, C.S., A bibliography on radial basis function approximation, *Boundary Element Communications*, **7**(1996), 155-163.
3. Cmnphob, M.M., *Examples of Partial Differential Equations*, (Japanese version translated by M. Sayigo), Morikata Press, Tokyo, 1982.
4. Bialecki, R. and Nowak, A. J., Boundary value problems for non-linear materials and non-linear boundary conditions, *Applied Mathematics Modelling*, **5** (1981), 417-421.
5. Chen, W. and Tanaka, M., Boundary knot method: A meshless, exponential convergence, integration-free, and boundary-only RBF technique, (submitted), 2000.